\documentclass[journal]{IEEEtran}
\usepackage{mathrsfs}
\usepackage{tikz}
\usepackage{color}
\usepackage{amsmath}
\usepackage{amssymb}
\usepackage{amsthm}
\usepackage{graphicx}
\usepackage{extarrows}
\usepackage{times}
\usepackage{graphics,color}
\newtheorem{theorem}{Theorem}	

\newtheorem{proposition}{Proposition}
\newtheorem{corollary}{Corollary}

\newtheorem{remark}{Remark}
\newcommand{\onetom}{1,\cdots,m}
\newcommand{\oneton}{1,\cdots,n}

\graphicspath{{figs/}}

\newtheorem{mylem}{Lemma}
\newtheorem{myprop}{Proposition}
\newtheorem{mydef}{Definition}

\begin{document}
\title{
Synchronization, Consensus of Complex Networks and Lyapunov Function Approach}

\author{Tianping~Chen%,~\IEEEmembership{Senior~Member,~IEEE}% <-this % stops a space
\thanks{This work is  supported by the National Natural Sciences Foundation of China under Grant Nos. 61273211.}
\thanks{T. Chen was with the School of Computer
Sciences/Mathematics, Fudan University, Shanghai 200433, China (tchen@fudan.edu.cn).
}}

\maketitle

\begin{abstract}
In this paper, we focus on the topic Synchronization and consensus of Complex Networks and their relationships. It is revealed that two topics are closely relating to each other and all results given in \cite{Li1,Li2} and many other papers can be obtained by the results in \cite{Chen1,Chen2}. It is pointed out that QUAD condition plays important role in discussing synchronization and consensus.

\end{abstract}


In recent years, lots of papers discussing synchronization of complex networks and consensus of multi-agents. both developed in parallel ways. For example,
in \cite{Chen1}, following model was discussed
\begin{align}\label{model1}
\frac{d x^{i}(t)}{dt}=f(x_{i}(t))
+c\sum\limits_{j=1}^{N}l_{ij}\Gamma x_{j}(t),\quad i=1,\cdots,N
\end{align}
where $x_{i}(t)\in R^{n}$ is the state variable of the $i-th$
node, $t\in [0,+\infty)$ is a continuous time,
$f:R\times[0,+\infty)\rightarrow R^{n}$ is continuous map,
$L=(l_{ij})\in R^{N\times N}$ is the coupling matrix with zero-sum
rows and $l_{ij}\ge 0$, for  $i\ne j$,  which is determined by the
topological structure of the LCODEs, and $\Gamma\in R^{n\times n}$ is an inner coupling matrix. Some time, picking
$\Gamma=diag\{\gamma_{1},\gamma_{2},\cdots,\gamma_{n}\}$ with
$\gamma_{i}\ge 0$, for $i=\oneton$.

On the other hand, based on controllable and detectable theory for linear systems, in \cite{Li1,Li2}, authors discussed following consensus of multiagent systems and
synchronization of complex networks
\begin{eqnarray}
\dot{x}_{i}(t)=Ax_{i}(t)+Bu_{i}(t),~~y_{i}(t)=Cx_{i}(t)%=(A+BK)x_{i}+BK(v_{i}(t)-x_{i}(t))
\end{eqnarray}
where $x_{i}(t)\in R^{n}$ is the stat, $u_{i}(t)\in R^{p}$ is the
control input, and $y_{i}(t)\in R^{q}$ is the measured output. $A\in R^{n\times n}$, $B\in R^{n\times p}$, $C\in R^{q\times n}$. It is assumed that is stabilizable and detectable.

An observer-type consensus protocol
%\begin{eqnarray}
%\dot{v}_{i}(t)=(A+BK)v_{i}(t)+F(\sum_{j=1}^{N}Cl_{ij}v_{j}(t)-\zeta_{i}(t))
%\end{eqnarray}
is proposed, which can  be written as
\begin{equation}\label{Li}
\left\{ \begin{array}{ll}
\dot{v}_{i}(t)=(A+BK)v_{i}(t)+\sum_{j=1}^{N}FCl_{ij}(v_{j}(t)-x_{j}(t)) \\
\dot{x}_{i}(t)=Ax_{i}(t)+BKv_{i}(t)
\end{array}
\right.
\end{equation}
where $K\in R^{p\times n}$, $F\in R^{n\times q}$. And $F$.

Let $e_{i}(t)=v_{i}(t)-x_{i}(t)$, one can transfer (\ref{Li}) to
\begin{equation}\label{e}
\left\{ \begin{array}{ll}
\dot{e}_{i}(t)=Ae_{i}(t)
+FC\sum_{j=1}^{N}l_{ij}e_{j}(t) \\
\dot{x}_{i}(t)=(A+BK)x_{i}(t)+BKe_{i}(t)
\end{array}
\right.
\end{equation}

In case $A+BK$ is controllable, the synchronization of the system (\ref{Li}) transfers to the synchronization of the system
\begin{align}\label{e1}
\dot{e}_{i}(t)=Ae_{i}(t)
+FC\sum_{j=1}^{N}l_{ij}e_{j}(t)
\end{align}

By checking two models (\ref{e}) and (\ref{e1}), it is easy to see that they can be unified in a general framwork.  Let $\Gamma=FC$. Then system (\ref{e1}) is a special case of (\ref{model1}). In fact, consensus of multiagent systems of complex networks can be viewed as special cases of the synchronization of nonlinear systems.

Therefore, what we need to do is to discuss synchronization model (\ref{model1}). %main results in \cite{Li1} can be obtained from the results given in \cite{Chen1}.

In this note, we show how to apply Lyapunov function approach to the synchronization and consensus of multi-agents.

\section{Some basic concepts and background}

Let's recall some basic concepts.

Following Lemma can be found in \cite{Chen1} (see Lemma 1 in \cite{Chen1}).
\begin{mylem}\quad %lem 1
If $L$ is a coupling matrix with Rank(L)=N-1, then the following items are valid:
\begin{enumerate}
\item If $\lambda$ is an eigenvalue of $L$ and $\lambda\neq 0$,
then  $Re(\lambda)<0$;

\item $L$ has an eigenvalue $0$ with
multiplicity 1 and the right eigenvector $[1,1,\dots,1]^{\top}$;

\item  Suppose $\xi=[\xi_{1},\xi_{2},\cdots,\xi_{m}]^{\top}\in
R^{m}$ (without loss of generality, assume
$\sum\limits_{i=1}^{m}\xi_{i}=1$) is the left eigenvector of $A$
corresponding to eigenvalue $0$. Then,  $\xi_{i}\ge 0$ holds for
all $i=\onetom$; more precisely,

\item $L$ is irreducible if and only if  $\xi_{i}>0$ holds for all
$i=\onetom$;

\item $L$ is reducible if and only if for some $i$, $\xi_{i}=0$.
In such case, by suitable rearrangement, assume that
$\xi^{\top}=[\xi_{+}^{\top},\xi_{0}^{\top}]$, where
$\xi_{+}=[\xi_{1},\xi_{2},\cdots,\xi_{p}]^{\top}\in R^{p}$, with
all $\xi_{i}>0$, $i=1,\cdots,p$, and
$\xi_{0}=[\xi_{p+1},\xi_{p+2},\cdots,\xi_{N}]^{\top}\in R^{N-p}$
with all $\xi_{j}=0$, $p+1\le j\le N$. Then, $L$ can be rewritten
as $\left[\begin{array}{cc}L_{11}& L_{12}\\L_{21}&
L_{22}\end{array}\right]$  where $L_{11}\in R^{p,p}$ is
irreducible and $L_{12}=0$.
\end{enumerate}
\end{mylem}

\begin{mydef}\quad Transverse space
$\mathcal
L=\{x=[{x_{1}}^{\top},{x_{2}}^{\top},\cdots,{x_{m}}^{\top}]^{\top}:
x_{i}\in
R^{n},~i=1,\cdots,m,~and~\sum\limits_{k=1}^{m}\xi_{k}x_{k}=0~\}$.
For the case of $n=1$ define
$L=\{(z_{1},\cdots,z_{m})^{\top},~z_{i}\in R,~
\sum\limits_{i=1}^{m}\xi_{i}z_{i}=0\}$
%For one dimensional case, we define
%$L=\{v=[v_{1},v_{2},\cdots,v_{m}]^{\top}:~
%\sum\limits_{k=1}^{m}\xi_{k}v_{k}=0,~v_{i}\in R\}$.
\end{mydef}
For the convenience of later use, introduce the following
notations: $\bar{x}(t)=\sum\limits_{i=1}^{m}\xi_{i}x_{i}(t)$,
$\bar{X}(t)=[\bar{x}^{\top}(t),\cdots,\bar{x}^{\top}(t)]^{\top}\in
\mathcal S $, which can be regarded as a projection of
$x(t)=[x^{\top}_{1}(t),\cdots,x^{\top}_{m}(t)]^{\top}$ on the
synchronization manifold $\mathcal S$ (generally,
nonorthogonal). %Under some conditions, we
Denote
$\delta{x}(t)=[\delta{x}_{1}(t)^{\top},\cdots,\delta{x}_{m}(t)^{\top}]^{\top}$,
where $\delta{x}_{i}(t)=x_{i}(t)-\bar{x}(t)$, $i=\onetom$. It is
easy to see that $\sum\limits_{i=1}^{m}\xi_{i}\delta{x}_{i}(t)=0$.

Thus, we have following result.
\begin{myprop}\quad
For any $x=(x^{\top}_{1},\cdots,x^{\top}_{m})^{\top}\in R^{mn}$,
we have $x=\bar{X}+\delta x$, where $\bar{X}$ and $\delta x$ are
defined as above, and it holds that $\bar{X}\in\mathcal S$ and
$\delta x\in\mathcal L$.
\end{myprop}

With this decomposition, the stability of the synchronization
manifold $\mathcal S$ for the model (\ref{model1}) is equivalent to $\delta {x}(t)\rightarrow
0$. Equivalently, the dynamical flow in the $(m-1)\times n$
dimensional subspace $\mathcal L$ converges to zero. In the
sequel, instead of investigating $x_{i}(t)$, we investigate
dynamical behaviors of $\delta{x}_{i}(t)$ directly.

Following function class  plays key role in discussing synchronization and consensus with Lyapunov functions.

\begin{mydef} \cite{book}
Function class $QUAD(\Delta,P)$: let
$P$ be a positive definite
matrix and $\Delta$ is any function.
$QUAD(\Delta,P)$ denotes a class of continuous
functions $f(x,t): R^{n}\times [0,+\infty)\rightarrow R^{n}$
satisfying
\begin{align}\label{QUADG}
(x-y)^{T}&P\bigg\{[f(x,t)-f(y,t)]-\Delta [x- y]\bigg\}\nonumber\\&\le -\epsilon
(x-y)^{T}(x-y)
\end{align}
for some $\epsilon>0$, all $x,y\in R^{n}$~and~$t>0$,
\end{mydef}

\begin{remark}
It is reasonable (or necessary) to assume that $P\Delta$ is positive definite in the range  $Ran(P\Delta)$. It is to ensure that $(x-y)^{T}P\Delta (x- y)>0$.
\end{remark}

\begin{remark}
In discussing synchronization of the model (\ref{e}), an important step is to compare
$(x-y)^{T}P\Delta (x- y)>0$ and $(x-y)^{T}P\Gamma (x- y)>0$.
\end{remark}

\begin{remark}
The concept was first introduced in \cite{Chen1} for the case that $P=diag\{p_{1},\cdots,p_{n}\}$ be a positive definite diagonal
matrix and $\Delta=diag\{\delta_{1},\cdots,\delta_{n}\}$ is a
diagonal matrix.
%\begin{align}\label{QUADO}
%(x-y)^{T}&P\bigg\{[f(x)-f(y)]-\Delta [x- y]\bigg\}\nonumber
%\\&\le -\epsilon
%(x-y)^{T}(x-y)
%\end{align}
%where $P$ is a positive definite matrix and $\Delta$ is any matrix, and $P\Delta$ should be positive definite in $Ran(P\Delta)$, where $Ran(A)$ denotes the range of the matrix  $A$. In some cases,  by a suitable coordinate transform, it can be seen that both (\ref{QUADO}) and (\ref{QUADG}) are equivalent.
\end{remark}

Let $P=Q^{T}JQ$, $\bar{x}=Qx$ be its eigenvalue decomposition, $\bar{f}(\bar{x})=Qf(x)=Qf(Q^{T}\bar{x})$, $\bar{\Delta}=Q\Delta$. Then, (\ref{QUADG}) can be written as
\begin{align*}%\label{QUADG}
(\bar{x}-\bar{y})^{T}&J\bigg\{[\bar{f}(\bar{x})-\bar{f}(\bar{y})]-\bar{\Delta} [\bar{x}-\bar{y}]\bigg\}\\
&\le-\epsilon(\bar{x}-\bar{y})^{T}(\bar{x}-\bar{y})
\end{align*}

If $\bar{\Delta}$ is also a positive diagonal matrix, then, the function $\bar{f}(\bar{x})$ satisfies the  QUAD condition introduced in \cite{Chen1}, where $J$ and $\bar{\Delta}$ are positive diagonal matrices.

In the following, we take (\ref{QUADG}) as QUAD definition.

\section{Synchronization Analysis of Complex Networks with Lyapunov functions}

Based on the synchronization state $\bar{x}(t)$, decomposition $\delta x(t)=x(t)-\bar{x}(t)$, and QUAD condition, synchronization problem of Complex Networks can be solved easily with Lyapunov function,
which was first proposed in \cite{Chen1}.

Since $\sum\limits_{j=1}^{m}l_{ij}=0$, it is
clear that
\begin{align*}
\sum\limits_{j=1}^{m}l_{ij}\Gamma
x^{j}(t)=\sum\limits_{j=1}^{m}l_{ij}\Gamma \delta{x}_{j}(t)
\end{align*}
Therefore,
\begin{align*}
\frac{d\delta {x}_{i}(t)}{dt}
&=
f(x^{i}(t))-f(\bar{x}(t))
+\sum\limits_{j=1}^{m}l_{ij}\Gamma\delta{x}_{j}(t)+J
\end{align*}
where
$J=f(\bar{x}(t))-\sum\limits_{k=1}^{m}\xi_{k}f(x_{k}(t))$ is independent of any index $i$.

Define a Lyapunov function as first proposed in \cite{Chen1}.
\begin{eqnarray*}
V(\delta{x})=\frac{1}{2}\delta{x}^{\top}\mathbf\Xi\mathbf
P\delta{x}=\frac{1}{2}\sum_{i=1}^{m}\xi_{i}\delta{x}_{i}^{\top}\mathbf
P\delta{x}_{i}
\end{eqnarray*}
where $\mathbf\Xi=\Xi\otimes I_{m}$ and $\mathbf P=I_{n}\otimes P$.

Denote $\delta y(t)=B^{T}\delta x(t)$, and differentiating $V(\delta{x})$ (noticing that
$\sum\limits_{i=1}^{m}\xi_{i}\delta{x}^{i}J=0$ and
$f(x)\in QUAD$ (\ref{QUADG}), we have
\begin{align*}
&\frac{d
V(\delta{x})}{dt}=\sum\limits_{i=1}^{m}\xi_{i}\delta{x}_{i}^{T}(t)P\frac{d
\delta x_{i}(t) }{dt}\nonumber\\
\le&-\epsilon \delta
x^{T}(t)\mathbf\Xi\delta
x(t) +\delta x^{T}(t)\bigg\{\Xi\otimes P\Delta+c(\Xi L)^{s}\otimes P\Gamma\bigg\}\delta x(t)
\end{align*}
If $\Xi\otimes P\Delta+c(\Xi L)^{s}\otimes P\Gamma$ is semi-negative definite in the transverse subspace $\mathcal L$ (notice $\delta x\in\mathcal L$), then
\begin{align*}
&\frac{d
V(\delta{x})}{dt}\le- \epsilon \delta
x^{T}(t)\mathbf\Xi\delta
x(t)
\end{align*}

We assume that $Ran(P\Delta)=Ran(P\Gamma)$, and $P\Gamma=BB^{T}$ is positive definite in the range $Ran(P\Gamma)$. Thus, for any $u\in R^{n}$,  there is a constant $c_{1}$, such that
\begin{align*}
u^{T}(P\Delta)u \le c_{1} u^{T}(P\Gamma)u
\end{align*}

%If $u\in Ran(P\Delta)\subseteq R^{n}$, then
%\begin{align*}
%u^{T}P\Delta u\le u^{T}||P\Delta||_{2}^{2} u
%\end{align*}
%and
%\begin{align*}
%u^{T}P\Delta u\le u^{T}[\Xi\otimes P\Delta]u\le %\max_{i=1,\cdots,m}\{\xi_{i}\}u^{T}||P\Delta||_{2}^{2} u
%\end{align*}

%\begin{align*}
%&\frac{d
%V(\delta{x})}{dt}\le- \epsilon \delta
%x^{T}(t)\mathbf\Xi\delta
%x(t)
%\end{align*}

Denote $\delta y(t)=B^{T}\delta x(t)$, then
\begin{align}
\delta x(t)^{T}[\Xi\otimes P\Delta] \delta x(t)\le c_{1}\max_{i=1,\cdots,m}\{\xi_{i}\}\delta y(t)^{T}\delta y(t)
\end{align}

On the other hand, let $0=\lambda_{1}>\lambda_{2}\ge\cdots\ge \lambda_{m}$ be the eigenvalues of the matrix $(\Xi A)^{s}$. Then

\begin{align*}
\delta x(t)^{T}[\{\Xi L\}^{s}\otimes P\Gamma] \delta x(t)&=\delta y(t)^{T}[\{\Xi L\}^{s}\otimes I_{n}]\delta y(t)\\
&\le \frac{\lambda_{2}}{\sqrt{m}\sqrt{\sum_{i=1}^{m} \xi_{i}^{2}}}\delta y(t)^{T}\delta y(t)
\end{align*}

%$$\delta x(t)^{T}[\{\Xi L\}^{s}\otimes P\Gamma] \delta x(t)\le \frac{\lambda_{2}}{\sqrt{m}\sqrt{\sum_{i=1}^{m} \xi_{i}^{2}}}\delta y(t)^{T}\delta y(t)$$
Therefore, in case that
\begin{align*}
 c>c_{1}\max_{i=1,\cdots,m}\{\xi_{i}\}(-\frac{\lambda_{2}}{\sqrt{m}\sqrt{\sum_{i=1}^{m} \xi_{i}^{2}}})^{-1}
\end{align*}
we have
\begin{align*}
 \delta x(t)^{T}[\Xi\otimes P\Delta] \delta x(t)+c \delta x(t)^{T}[\{\Xi L\}^{s}\otimes P\Gamma] \delta x(t)< 0
\end{align*}
and there exists a constant $\bar{c}>0$, such that
\begin{align*}
&\frac{d
V(\delta{x})}{dt}\le -\epsilon \delta
x^{T}(t)\mathbf\Xi\delta
x(t)<-\bar{c} V(\delta{x})
\end{align*}
and $V(t)$ converges to zero exponentially.

%The reader also can refer to \cite{book}.

Now, we can give following
\begin{proposition}
Under the QUAD condition (\ref{QUADG}), $Ran(P\Delta)= Ran(P\Gamma)$, where $Ran(A)$ denotes the range of the matrix  $A$, (in particular, $\Delta=\Gamma$). and $P\Gamma=BB^{T}$ is positive definite in the range $Ran(P\Gamma)$.
The system (\ref{model1}) can reach synchronization if the coupling strength $c$ is large enough.
\end{proposition}

\subsection{Pinning Control Synchronization of Complex Networks}

Let $s(t)$ is a solution of $\dot{s}(t)=g(s(t))$.

Consider the following pinning control model
\begin{eqnarray}\label{pin}
\left\{\begin{array}{cc}\frac{dx_1(t)}{dt}&=f(x_1(t))
+c\sum\limits_{j=1}^m l_{1j}\Gamma x_j(t)\\
&-c\varepsilon(x_{1}(t)-s(t)),\\
\frac{dx_i(t)}{dt}&=g(x_i(t))+c\sum\limits_{j=1}^ml_{ij}\Gamma x_j(t),\\
&i=2,\cdots,m\end{array}\right.
\end{eqnarray}

As addressed in \cite{Chen2}, following proposition plays key role.

\begin{proposition} (see \cite{Chen2}) If  $L=(l_{ij})_{i,j=1}^{m}$ is an
irreducible matrix with $Rank(L)=m-1$, satisfying
$l_{ij}\geq0,$  if $ i\not=j$, and
$\sum\limits_{j=1}^{m}l_{ij}=0$, for $i=1,2,\cdots,m$.
\[\tilde{L}=\left(\begin{array}{cccc}l_{11}-\varepsilon&l_{12}&\cdots&l_{1m}\\
l_{21}&l_{22}&\cdots&l_{2m}\\\vdots&\vdots&\ddots&\vdots\\
l_{m1}&l_{m2}&\cdots&l_{mm}\end{array}\right)\]
$\xi=[\xi_{1},\cdots,\xi_{m}]^{T}$ is the eigenvector of $L$ with eigenvalue $0$. Then, $\tilde{L}$ is a non-singular M-matrix; all the eigenvalues of $\tilde{L}$ have negative real part, and all the eigenvalues of the matrix $\{\Xi \tilde{L}\}^{s}=\frac{1}{2}[\Xi \tilde{L}+\tilde{L}^{T}\Xi]$, where $\Xi=diag[\xi_{1},\cdots,\xi_{m}]$, are negative.
\end{proposition}
Then, (\ref{pin}) can be rewritten as
\begin{equation}\label{pina}
\dot{x}_{i}(t)=f(x_{i}(t))
+c\sum_{j=1}^{N}\tilde{l}_{ij}\Gamma x_{j}(t),~~i=1,\cdots,m
\end{equation}

Now, replacing $\bar{x}(t)$ by $s(t)$, let $\tilde\delta {x}_{j}(t)=x_{j}(t)-s(t)$, and define a Lyapunov function as
\begin{eqnarray*}
V_{1}(\tilde\delta{x})=\frac{1}{2}\delta{\tilde{x}}^{T}\mathbf\Xi\mathbf
P\tilde\delta{x}=\frac{1}{2}\sum_{i=1}^{m}\xi_{i}\tilde\delta{x}_{i}^{T}\mathbf
P\tilde\delta{x}_{i}
\end{eqnarray*}

Similar to previous arguments, by the QUAD condition (\ref{QUADG}), we have:
\begin{align*}
&\frac{d
V_{1}(\tilde\delta{x})}{dt}=\sum\limits_{i=1}^{m}\xi_{i}\tilde\delta{x}_{i}^{T}(t)P\frac{d
\tilde\delta{x}_{i}(t) }{dt}\nonumber\\
\le&-\epsilon \tilde\delta{x}^{T}(t)\mathbf\Xi\tilde\delta{x}(t) +\delta
y^{T}(t)[\mathbf\Xi+c(\Xi \tilde{L})^{s}\otimes I_{n}]\delta
y(t)\end{align*}
where $\delta y(t)=B^{T}\tilde\delta{x}(t)$.

Let $0>\mu_{1}\ge \mu_{2}\ge\cdots\ge \mu_{m}$ be the eigenvalues of $(\Xi \tilde{L})^{s}$, and $c>\frac{\max_{i=1,\cdots,m}\{\xi_{i}\}}{|\mu_{1}|}$, then
there exists a constant $c_{1}>0$, such that
\begin{align*}
\delta
y^{T}(t)[\mathbf\Xi+c(\Xi \tilde{L})^{s}\otimes I_{n}]\delta
y(t)<0
\end{align*}
which implies
\begin{align*}
&\frac{d
V_{1}(\tilde\delta{x})}{dt}\le-\epsilon \tilde\delta{x}^{T}(t)\mathbf\Xi\tilde\delta{x}(t)<-c_{1} V_{1}(\tilde\delta{x})
\end{align*}
and $V_{1}(t)$ converges to zero exponentially.

\begin{proposition}
Under the QUAD condition (\ref{QUADG}), $Ran(P\Delta)=Ran(P\Gamma)$ (in particular, $\Delta=\Gamma$). and $P\Gamma=BB^{T}$ is positive definite in the range $Ran(P\Gamma)$, the system (\ref{pin}) can synchronize all $x_{i}$ to $s(t)$, if the coupling strength is large enough.
\end{proposition}

\begin{remark}
In some paper, authors discussed leader-follower system. In fact, it is just a special case of pinning control system with one controller. It was clearly addressed in \cite{Chen2}.
\end{remark}

\begin{remark}
By choosing various $P$, $\Gamma$, $\Delta$, we can give different criteria. For example,
in \cite{Yu1,Yu2}, authors discussed the pinning control system
\begin{equation}%\label{pina}
\dot{x}_{i}(t)=f(x_{i}(t))
+c\sum_{j=1}^{N}\tilde{l}_{ij}\Gamma x_{j}(t),~~i=1,\cdots,m
\end{equation}
where function $f(x)$ satisfies following condition
\begin{align}%\label{QUADG}
(x-y)^{T}&\bigg\{[f(x)-f(y)]-K\Gamma [x- y]\bigg\}\le 0
\end{align}
and $\Gamma$ and $K\Gamma $ are assumed positive definite and commutable.

%$$\theta I_{m}+c(G-D)<0$$
They claimed that in case that
$$\theta I_{m}+c\tilde{L}<0$$
where $\theta=||K||$, then, all $x_{i}(t)$ synchronize to $s(t)$.

for this case, we first prove that if two matrices $\Gamma$ and $K$ are commutable, then $\Gamma$ and $K$ have same eigenvectors. In fact, if $x$ is an eigenvector of $\Gamma$ with eigenvalue $\lambda$, then
$$\Gamma x=\lambda x$$
Because $\Gamma K=K\Gamma$, then
$$\Gamma K x=K\Gamma x=\lambda K x$$
which means that $Kx$ is also an eigenvector of the matrix $\Gamma$ with eigenvalue $\lambda$, and there exists a constant $\mu$ such that
$$Kx=\mu x$$
which means that $\Gamma$ and $K$ have same eigenvectors.

Therefore, if $\Gamma=Q\gamma Q^{T}$, where  $\gamma=diag\{\gamma_{1},\cdots,\gamma_{n}\}$ are the eigenvalues of $\Gamma$, and $K=Qk Q^{T}$, where  $k=diag\{k_{1},\cdots,k_{n}\}$ are the eigenvalues of $\Gamma$, then $K\Gamma=Qk\gamma Q^{T}$.

let $P=I_{n}$, $\Delta=K\Gamma$, we have following QUAD condition
\begin{align}\label{QUADG1}
(x-y)^{T}\bigg\{[f(x,t)-f(y,t)]-\Delta [x-y]\bigg\}\le 0
\end{align}

Define
\begin{eqnarray*}
V_{1a}(\tilde\delta{x})=\frac{1}{2}\delta{\tilde{x}}^{T}\mathbf\Xi\tilde\delta{x}
=\frac{1}{2}\sum_{i=1}^{m}\xi_{i}\tilde\delta{x}_{i}^{T}
\tilde\delta{x}_{i}
\end{eqnarray*}

\begin{align*}
&\frac{d
V_{1a}(\tilde\delta{x})}{dt}=\sum\limits_{i=1}^{m}\xi_{i}\tilde\delta{x}_{i}^{T}(t)\frac{d
\tilde\delta{x}_{i}(t) }{dt}\nonumber\\
=&\tilde\delta{x}^{T}(t)[\Xi\otimes K\Gamma+c(\Xi \tilde{L})^{s}\otimes \Gamma]\tilde\delta{x}(t)
\end{align*}
Let $\theta=\max_{i=1,\cdots,n,j=1,\cdots,m}\{\xi_{i}k_{j}\gamma_{j}\}$.
Then,
\begin{align}
\tilde\delta{x}^{T}(t)[\Xi\otimes K\Gamma]\tilde\delta{x}(t)
\le \theta
\tilde\delta{x}^{T}(t)\tilde\delta{x}(t)
\end{align}

Similar to previous arguments, let $0>\mu_{1}\ge \mu_{2}\ge\cdots\ge \mu_{m}$ be the eigenvalues of $(\Xi \tilde{L})^{s}$, and $c>\frac{\theta}{|\mu_{1}|}$, then
\begin{align*}
&\frac{d
V_{1a}(\tilde\delta{x})}{dt}\le-(c-\frac{\theta}{|\mu_{1}|}) \tilde\delta{x}^{T}(t)\mathbf\Xi\tilde\delta{x}(t)<-c_{1} V_{1a}(\tilde\delta{x})
\end{align*}
and $V_{1}(t)$ converges to zero exponentially.

It is clear that the results given in \cite{Yu1,Yu2} can be obtained as a direct consequences of the derivation given above. Moreover, the result is much better.

\end{remark}
\subsection{Adaptive Algorithms}

In previous parts, we revealed that we can always synchronize of pinning
a coupled complex network if
the coupling strength is large enough. However, in practice, it
is not allowed that the coupling strength is arbitrarily large.
For synchronization, it was pointed out in \cite{Chen3} that theoretical
value of the coupling strength is much larger than needed
in practice. Therefore, the following question was arisen in
\cite{Chen3}: Can we find the sharp bound cmin? Similarly, in pinning
process, it is also important to make the coupling strength as
small as possible. It is clear that theoretical value of strength
given in previous theorems are heavily based on the QUAD condition, which is too strong.
Therefore, it is possible to lessen coupling strength dramatically.

1. {Adaptive Algorithms for synchronization}

For this purpose, consider following adaptive algorithm
\begin{eqnarray}
\left\{\begin{array}{cc}
\frac{dx_i(t)}{dt}&=f(x_i(t))+c(t)\sum\limits_{j=1}^ml_{ij}\Gamma x_j(t),\\
 &\dot{c}(t) =\frac{\alpha}{2}\sum\limits_{i=1}^m\delta
x_i^T(t)BB^{T}\delta x_i(t)\end{array}\right. \label{adap1}
\end{eqnarray}
where $\Gamma=P^{-1}BB^{T}$ $\delta x_i(t)=x_i(t)-\bar{x}(t)$.
%where $\delta x_i(t)=x_i(t)-\bar{x}(t)$ for synchronization/consensus without control.
%\begin{eqnarray}
%\left\{\begin{array}{cc}
%&\dot{c}(t) =\frac{\alpha}{2}\sum\limits_{i=1}^m\delta
%x_i^T(t)BB^{T}\delta x_i(t)\end{array}\right. \label{adap2}
%\end{eqnarray}
%$\delta \tilde{x}_i(t)=x_i(t)-s(t)$.
%for synchronization/consensus with pinning control, which was discussed in \cite{Chen2}.

%Here, we just give a brief proof for the adaptive algorithm (\ref{adap2}), which can be %traced to \cite{Chen2}.

Pick a constant $\alpha>0$. Define a Lyapunov
function
\begin{eqnarray}
V_{2}(\delta{x}(t))=\frac{1}{2}\sum\limits_{i=1}^m\xi_i\delta x_i(t)P\delta
x_i^T(t)+\frac{\beta}{\alpha}(c-c(t))^2;
\end{eqnarray}
where constants $c$ and $\beta$ will be decided later.

Differentiating $V_{2}(\delta{x})$ (noticing that
$\sum\limits_{i=1}^{m}\xi_{i}\delta{x}^{i}J=0$ and
$f(x)\in QUAD$ (\ref{QUADG}), we have
\begin{align*}
&\frac{d
V_{2}(\delta{x}(t))}{dt}=\sum\limits_{i=1}^{m}\xi_{i}\delta{x}_{i}^{T}(t)P\frac{d
\delta x_{i}(t) }{dt}\nonumber\\
\le&-\epsilon \sum_{i=1}^{m}\xi_{i}\delta{x}_{i}^{T}(t)\delta
x_{i}(t)+\sum_{i=1}^{m}\xi_{i}\delta{x}_{i}^{T}(t)BB^{T}\delta
x_{i}(t)\\
&+c(t)\sum_{i,j=1}^{m}\xi_{i}\delta{x}_{i}^{T}(t)l_{ij}BB^{T}\delta
x_{j}(t)\\
&-\beta c\sum_{i=1}^m\xi_{i}\delta x_i^T(t)BB^{T}\delta
x_i(t)\\&+\beta c(t)\sum_{i=1}^m \xi_{i}\delta x_i^T(t)BB^{T}\delta x_i(t)
\end{align*}

It is easy to check that
\begin{align*}
\sum_{i=1}^m \xi_{i}\delta x_i^T(t)BB^{T}\delta x_i(t)\le \max_{i=1,\cdots,m}(\xi_{i})\sum_{i,j=1}^{m}\delta{x}_{i}^{T}(t)BB^{T}\delta
x_{j}(t)
\end{align*}
and
\begin{align*}
\sum_{i,j=1}^{m}\xi_{i}\delta{x}_{i}^{T}(t)l_{ij}BB^{T}\delta
x_{j}(t)\le \lambda_{2}\sum_{i,j=1}^{m}\delta{x}_{i}^{T}(t)BB^{T}\delta
x_{j}(t)
\end{align*}
Therefore, in case that $\lambda_{2}+\beta\max_{i=1,\cdots,m}(\xi_{i})<0$, we have
\begin{align*}
&c(t)\sum_{i,j=1}^{m}\xi_{i}\delta{x}_{i}^{T}(t)l_{ij}BB^{T}\delta
x_{j}(t)+\beta c(t)\sum_{i=1}^m \xi_{i}\delta x_i^T(t)BB^{T}\delta x_i(t)\\
&\le c(t)(\lambda_{2}+\beta\max_{i=1,\cdots,m}(\xi_{i}))\sum_{i,j=1}^{m}\delta{x}_{i}^{T}(t)BB^{T}\delta
x_{j}(t) \le 0
\end{align*}
%if $\lambda_{2}+\beta\max_{i=1,\cdots,m}(\xi_{i})<0$. %$\beta<\frac{-\lambda_{2}}{\max_{i=1,\cdots,m}(\xi_{i})}$

Furthermore, pick $c>\frac{1}{\beta}$, we have
\begin{align*}
&\frac{d
V(\delta{x})}{dt}\le-\epsilon \sum_{i=1}^{m}\xi_{i}\delta{x}_{i}^{T}(t)\delta
x_{i}(t)
\end{align*}
Therefore,
\begin{eqnarray*}
\epsilon\int_{0}^{\infty}\sum\limits_{i=1}^m\xi_{i}\delta x_i(t)^{T}\delta
x_i(t)dt<V_{2}(0)-V_{2}(\infty)
\end{eqnarray*}
which implies $\delta x_i(t)\rightarrow 0$ and $c(t)\rightarrow c_0$, where
$c_0$ is a positive constant. The proof is completed.

Therefore, we have
\begin{proposition}
Suppose $L$ is a connecting coupling matrix and QUAD condition (\ref{QUADG}) is satisfied,  $Ran(P\Delta)\subseteq Ran(P\Gamma)$, (in particular, $\Delta=\Gamma$). and $P\Gamma=BB^{T}$ is positive definite in the range $Ran(P\Gamma)$, if the coupling strength is large enough.  then the adaptive algorithm (\ref{adap1}) can reach synchronization.
\end{proposition}

\begin{remark}
all the results apply to adaptive synchronization model with pinning control
\begin{eqnarray}
\left\{\begin{array}{cc}
\frac{dx_i(t)}{dt}&=f(x_i(t))+c(t)\sum\limits_{j=1}^m\tilde{l}_{ij}\Gamma x_j(t),\\
 &\dot{c}(t) =\frac{\alpha}{2}\sum\limits_{i=1}^m\delta
x_i^T(t)BB^{T}\delta x_i(t)\end{array}\right. \label{adap1}
\end{eqnarray}
where $\Gamma=P^{-1}BB^{T}$ $\delta x_i(t)=x_i(t)-s(t)$. Details are omitted.
\end{remark}

2. {Distributive adaptive Algorithms for synchronization with pinning control}

Consider the following distributive adaptive pinning control model
\begin{eqnarray}%\label{pin}
\left\{\begin{array}{cc}\frac{dx_1(t)}{dt}&=f(x_1(t))
+c\sum\limits_{j=1}^m l_{1j}\Gamma x_j(t)\\
&-c\varepsilon(x_{1}(t)-s(t)),\\
\frac{dx_i(t)}{dt}&=g(x_i(t))+c\sum\limits_{j=1}^ml_{ij}\Gamma x_j(t),\\
&i=2,\cdots,m
\\
 &\dot{c}_{i}(t) =\frac{\alpha}{2}\tilde\delta
x_i^T(t)BB^{T}\tilde\delta x_i(t)
\end{array}\right.
\end{eqnarray}
where $\Gamma=P^{-1}BB^{T}$ $\tilde\delta x_i(t)=x_i(t)-s(t)$.

It can also be written as
\begin{eqnarray}
\left\{\begin{array}{cc}\frac{dx_i(t)}{dt}&=f(x_i(t))
+\sum\limits_{j=1}^m c_{i}(t)\tilde{l}_{1j}P^{-1}B^{T}Bc_{j}(t)x_j(t),\\
 &\dot{c}_{i}(t) =\frac{\alpha}{2}\tilde\delta
x_i^T(t)BB^{T}\tilde\delta x_i(t)\end{array}\right. \label{adap3}
\end{eqnarray}

Define a Lyapunov function as
\begin{eqnarray}
V_{2a}(\tilde\delta{x}(t))=\frac{1}{2}\sum\limits_{i=1}^m\xi_i\tilde\delta x_i(t)P\tilde\delta
x_i^T(t)+\frac{\beta}{\alpha}\sum\limits_{i=1}^m(c-c_{i}(t))^2
\end{eqnarray}
where constants $c$ and $\beta$ will be decided later.

Differentiating $V_{2a}(\tilde\delta{x})$ (noticing that
$f(x)\in QUAD$ (\ref{QUADG}), we have
\begin{align*}
&\frac{d
V_{2a}(\tilde\delta{x}(t))}{dt}=\sum\limits_{i=1}^{m}\xi_{i}\tilde\delta{x}_{i}^{T}(t)P
\frac{d\tilde\delta x_{i}(t) }{dt}\nonumber\\
\le&-\epsilon \sum_{i=1}^{m}\xi_{i}\tilde\delta{x}_{i}^{T}(t)\tilde\delta
x_{i}(t)+\sum_{i=1}^{m}\xi_{i}\tilde\delta{x}_{i}^{T}(t)BB^{T}\tilde\delta
x_{i}(t)\\
&+\sum_{i,j=1}^{m}\xi_{i}c_{i}(t)\tilde\delta{x}_{i}^{T}(t)\tilde{l}_{ij}BB^{T}c_{j}(t)
\tilde\delta x_{j}(t)\\
&-\beta c\sum_{i=1}^m\xi_{i}\tilde\delta x_i^T(t)BB^{T}\tilde\delta
x_i(t)\\&+\beta \sum_{i=1}^m c_{i}(t)\xi_{i}\tilde\delta x_i^T(t)BB^{T}\tilde\delta x_i(t)
\end{align*}
Let $c_{1}=\min_{i=1,\cdots,m}c_{i}(0)$, $c_{2}=\max_{i=1,\cdots,m}\xi_{i}$, then
\begin{align*}
&\sum_{i=1}^m \xi_{i}c_{i}(t)\tilde\delta x_i^T(t)BB^{T}\tilde\delta x_i(t)\\
&\le\frac{c_{2}}{c_{1}}\sum_{i=1}^m c_{i}(t)\tilde\delta x_i^T(t)BB^{T}c_{i}(t)\tilde\delta x_i(t)
\end{align*}

Let $0>\mu_{1}\ge \mu_{2}\ge\cdots\ge \mu_{m}$ be the eigenvalues of $(\Xi \tilde{L})^{s}$, and $c>\frac{\max_{i=1,\cdots,m}\{\xi_{i}\}}{|\mu_{1}|}$,
\begin{align*}
&\sum_{i,j=1}^{m}\xi_{i}c_{i}(t)\tilde\delta{x}_{i}^{T}(t)\tilde{l}_{ij}BB^{T}c_{j}(t)\tilde\delta
x_{j}(t)\\
&\le \mu_{1}\sum_{i=1}^m c_{i}(t)\tilde\delta x_i^T(t)BB^{T}c_{i}(t)\tilde\delta x_i(t)
\end{align*}
Pick $\beta<\frac{c_{1}}{c_{2}}|\mu_{1}|$ sufficiently small, we have
\begin{align*}
&\beta \sum_{i=1}^m \xi_{i}c_{i}(t)\tilde\delta x_i^T(t)BB^{T}\tilde\delta x_i(t)\\
&+\sum_{i,j=1}^{m}\xi_{i}c_{i}(t)\tilde\delta{x}_{i}^{T}(t)\tilde{l}_{ij}BB^{T}c_{j}(t)\tilde\delta
x_{j}(t)<0
\end{align*}
Then, pick $c$ sufficiently large such that $\beta c>1$, we have
\begin{align*}
\sum_{i=1}^{m}\xi_{i}\tilde\delta{x}_{i}^{T}(t)BB^{T}\tilde\delta
x_{i}(t)<-\beta c\sum_{i=1}^m\xi_{i}\delta x_i^T(t)BB^{T}\delta
x_i(t)
\end{align*}
which implies
\begin{align*}
&\frac{d
V(\delta{x})}{dt}\le-\epsilon \sum_{i=1}^{m}\xi_{i}\tilde\delta{x}_{i}^{T}(t)\tilde\delta
x_{i}(t)
\end{align*}
Therefore,
\begin{eqnarray*}
\epsilon\int_{0}^{\infty}\sum\limits_{i=1}^m\xi_{i}\delta x_i(t)^{T}\delta
x_i(t)dt<V_{2}(0)-V_{2}(\infty)
\end{eqnarray*}
which implies $\delta x_i(t)\rightarrow 0$ and $c_{i}(t)\rightarrow c_i$, where
$c_i$ are positive constants. Therefore, we have
\begin{proposition}
Suppose $L$ is a connecting coupling matrix and QUAD condition (\ref{QUADG}) is satisfied,  $Ran(P\Delta)\subseteq Ran(P\Gamma)$, (in particular, $\Delta=\Gamma$). and $P\Gamma=BB^{T}$ is positive definite in the range $Ran(P\Gamma)$, if the coupling strength is large enough.  then the distributive adaptive algorithm (\ref{adap3}) can reach pinning synchronization.
\end{proposition}

3. {Adaptive Algorithms for synchronization by adapting coupling weights}

Now, we discuss folowing adaptive algorithm
\begin{align}
\left\{\begin{array}{ll}\frac{dx_i(t)}{dt}&=f(x_i(t)))+c\sum\limits_{j=1}^m w_{ij}(t)\Gamma(x_j(t)-x_{i}(t)),\\
 &\dot{w}_{ij}(t) =\rho_{ij}\xi_{i}
[x_i(t)-\bar{x}(t)]^{T}\Gamma[x_i(t)-x_j(t)]
\end{array}\right. \label{adap4}
\end{align}
proposed in \cite{Chen4}, which adapts all weights for synchronization.
\begin{align}
\dot{x}_{i}(t)=f(x_i(t))+c\sum_{j=1}^{m}l_{ij}\Gamma x_{j}(t)
\end{align}
 readers can refer to \cite{Chen5}.

Define Lyapunov function
\begin{align}
V_{3}(t)=&\frac{1}{2}\sum_{i=1}^{m}\xi_{i}(x_i(t)-\bar{x}(t))^{T}P(x_i(t)-\bar{x}(t))\nonumber\\
&+\frac{1}{2}\sum_{i=1}^{m}\sum_{j=1}^{m}\frac{1}{\rho_{ij}}(w_{ij}(t)-cl_{ij})^2
\end{align}
\begin{align}
&\dot{V}_{3}(t)
=\sum_{i=1}^{m}\xi_{i}(x_i(t)-\bar{x}(t))^{T}
P(f(x_{i}(t))-f(\bar{x}(t))\nonumber\\
&+\sum_{i,j=1}^{m}\xi_{i}(x_i(t)-\bar{x}(t))^{T}w_{ij}(t)P\Gamma (x_{j}(t)-x_{i}(t))
\nonumber\\
&+\sum_{i,j=1}^{m}(w_{ij}(t)-cl_{ij})\xi_{i}(x_{i}(t)-\bar{x}(t))^{T}
P\Gamma(x_{i}(t)-x_{j}(t))\nonumber\\
&=\sum_{i=1}^{m}\xi_{i}(x_i(t)-\bar{x}(t))^{T}P(f(x_{i}(t))-f(\bar{x}(t)))\nonumber\\
&+c\sum_{i=1}^{m}\sum_{j=1}^{m}\xi_{i}(x_i(t)-\bar{x}(t))^{T}l_{ij}P\Gamma(x_{j}(t)-x_{i}(t))
\nonumber\\
&=\sum_{i=1}^{m}\xi_{i}(x_i(t)-\bar{x}(t))^{T}P(f(x_{i}(t))-f(\bar{x}(t)))\nonumber\\
&+c\sum_{i=1}^{m}\sum_{j=1}^{m}\xi_{i}(x_i(t)-\bar{x}(t))^{T}l_{ij}P\Gamma(x_{j}(t)-\bar{x}(t))
\end{align}
By similar arguments, for sufficient large $c$, we have
\begin{eqnarray*}
\dot{V}_{3}(t)<-\epsilon\xi_{i}\sum\limits_{i=1}^m\xi_{i}\delta x_i(t)^{T}\delta
x_i(t)
\end{eqnarray*}
and
\begin{eqnarray*}
\epsilon\int_{0}^{\infty}\sum\limits_{i=1}^m\xi_{i}\delta x_i(t)^{T}\delta
x_i(t)<V(0)-V(\infty)
\end{eqnarray*}
which implies $\delta x(t)\rightarrow 0$.

Therefore, we can give
\begin{proposition}
Suppose $L$ is a connecting coupling matrix and QUAD condition (\ref{QUADG}) is satisfied,  $Ran(P\Delta)\subseteq Ran(P\Gamma)$, (in particular, $\Delta=\Gamma$). and $P\Gamma=BB^{T}$ is positive definite in the range $Ran(P\Gamma)$, if the coupling strength is large enough.  then the adaptive algorithm (\ref{adap4}) by adapting coupling weights can reach synchronization.
\end{proposition}

3. {Distributive adaptive Algorithms for synchronization by adapting coupling weights}

In case the coupling matrix is symmetric, one can use the so called distributive adaptive algorithm
\begin{align}\label{Li3adap}
\left\{\begin{array}{ll}\frac{dx_i(t)}{dt}&=f(x_i(t))+c\sum\limits_{j=1}^m w_{ij}(t)\Gamma(x_j(t)-x_{i}(t)),\\
\dot{w}_{ij}(t) &=\rho_{ij}
[x_i(t)-x_j(t)]^{T}\Gamma[x_i(t)-x_j(t)]
\end{array}\right.
\end{align}
for the system
\begin{align}
\dot{x}_{i}(t)=f(x_i(t))+c\sum_{j=1}^{m}l_{ij}\Gamma x_{j}(t)
\end{align}

In this case, we need following identity for a symmetric coupling matrix $L=[l_{ij}]$
\begin{align}
&\sum_{i=1}^{m}x_i^{T}(t)l_{ij} y_j(t)\nonumber
\\&=-\sum_{i>j}l_{ij}(x_{j}(t)-x_{i}(t))^{T}(y_{j}(t)-y_{i}(t))
\end{align}
Therefore,
\begin{align}\label{sym}
&\sum_{i=1}^{m}l_{ij}(x_i(t)-\bar{x}(t))^{T}P(x_j(t)-\bar{x}(t))\nonumber
\\&=-\sum_{i>j}l_{ij}(x_{j}(t)-x_{i}(t))^{T}P(x_{j}(t)-x_{i}(t))
\end{align}

Denote $\bar{x}(t)=\frac{1}{m}\sum_{i=1}^{m}x_{i}(t)$ and the Lyapunov function
\begin{align}
V_{3a}(t)=&\frac{1}{2}\sum_{i=1}^{m}(x_i(t)-\bar{x}(t))^{T}P(x_i(t)-\bar{x}(t))\nonumber\\
&+\frac{1}{2}\sum_{i=1}^{m}\sum_{j=1}^{m}\frac{1}{\rho_{ij}}(w_{ij}(t)-cl_{ij})^2
\end{align}
Differentiating it, we have
\begin{align}
&\dot{V}_{3a}(t)
=\sum_{i=1}^{m}(x_i(t)-\bar{x}(t))^{T}
P(f(x_{i}(t))-f(\bar{x}(t))\nonumber\\
&+\sum_{i,j=1}^{m}(x_i(t)-\bar{x}(t))^{T}w_{ij}(t)P\Gamma (x_{j}(t)-\bar{x}(t))
\nonumber\\
&+\sum_{i,j=1}^{m}(w_{ij}(t)-cl_{ij})(x_{i}(t)-x_{j}(t))^{T}
P\Gamma(x_{i}(t)-x_{j}(t))
\end{align}
By the assumption that $w_{ij}(t)$ and $(l_{ij})$ are symmetric and noticing (\ref{sym}), we have
\begin{align}
&\dot{V}_{3a}(t)
=\sum_{i=1}^{m}(x_i(t)-\bar{x}(t))^{T}
P(f(x_{i}(t))-f(\bar{x}(t))\nonumber\\
&+\sum_{i,j=1}^{m}(x_i(t)-x_{j}(t))^{T}w_{ij}(t)P\Gamma (x_{j}(t)-x_i(t))
\nonumber\\
&+\sum_{i,j=1}^{m}(w_{ij}(t)-cl_{ij})(x_{i}(t)-x_{j}(t))^{T}
P\Gamma(x_{i}(t)-x_{j}(t))\nonumber\\
&=\sum_{i=1}^{m}(x_i(t)-\bar{x}(t))^{T}P(f(x_{i}(t))-f(\bar{x}(t)))\nonumber\\
&+c\sum_{i=1}^{m}\sum_{j=1}^{m}(x_i(t)-x_{j}(t))^{T}l_{ij}P\Gamma(x_{j}(t)-x_{i}(t))
\nonumber\\
&=\sum_{i=1}^{m}(x_i(t)-\bar{x}(t))^{T}P(f(x_{i}(t))-f(\bar{x}(t)))\nonumber\\
&+c\sum_{i=1}^{m}\sum_{j=1}^{m}(x_i(t)-\bar{x}(t))^{T}l_{ij}P\Gamma(x_{j}(t)-\bar{x}(t))
\end{align}

By previous similar arguments, we have
\begin{proposition}
If $L$ is a \underline{symmetric} connecting coupling matrix and QUAD condition (\ref{QUADG}) is satisfied,  $Ran(P\Delta)\subseteq Ran(P\Gamma)$, (in particular, $\Delta=\Gamma$). and $P\Gamma=BB^{T}$ is positive definite in the range $Ran(P\Gamma)$, if the coupling strength is large enough.  then the adaptive algorithm (\ref{Li3adap}) by adapting coupling weights can reach synchronization.
\end{proposition}

\section{Consensus of Multiagent Systems of Complex Networks with Lyapunov functions}

In this part, as applications of the results obtained in previous section, we discuss Consensus of Multiagent Systems.

Recently, Consensus Protocols for Linear Multi-Agent Systems has attracted some researchers' attention see \cite{Li1,Li2}.

By letting $f(x_{i}(t))=Ax_{i}(t)$, $c\Gamma=FC$, it is easy to see that model (\ref{e1}) is a special case of the model (\ref{model1}).
%\begin{align}
%\frac{d x_{i}(t)}{dt}=Ax_{i}(t)
%+c\sum\limits_{j=1}^{N}l_{ij}\Gamma x_{j}(t),\quad i=1,\cdots,N
%\end{align}
Then, all results obtained in previous section can apply to the consensus

First of all, we discuss the case $(A,C)$ is detectable. In this case, for any fixed $t$, let
\begin{align*}
P(t)=2\int_{0}^{t}e^{-A^{T}t}C^{T}Ce^{-At}dt>0
\end{align*}
then
\begin{align*}
P(t)A+A^{T}P(t) &=-2\int_{0}^{t}\frac{d}{dt}[e^{-A^{T}s}C^{T}Ce^{-As}]ds\\
&=2C^{T}C-2e^{-A^{T}t}C^{T}Ce^{-At}
\end{align*}

Denote
$$P=\frac{2}{t_{2}-t_{1}}\int_{t_{1}}^{t_{2}}P(t)dt$$
then
\begin{align*}
PA+A^{T}P =2C^{T}C-\frac{2}{t_{2}-t_{1}}\int_{t_{1}}^{t_{2}}2e^{-A^{T}t}C^{T}Ce^{-At}dt
\end{align*}
Therefore, there exists $\epsilon>0$ such that
\begin{align}\label{LMIC}
PA+A^{T}P-2C^{T}C =&-\frac{2}{t_{2}-t_{1}}\int_{t_{1}}^{t_{2}}2e^{-A^{T}t}C^{T}Ce^{-At}dt
\nonumber\\<&-\epsilon I_{n}
\end{align}
which is equivalent to the QUAD condition
\begin{align}\label{QUADG1}
(x-y)^{T}P\bigg\{A(x-y)-\Delta[x- y]\bigg\}\le -\epsilon(x-y)^{T}(x-y)
\end{align}
where $\Delta=P^{-1}C^{T}C$.

Therefore, noticing $\Delta=\Gamma$, by Proposition 1, we have
\begin{proposition}
Under the QUAD condition (\ref{QUADG1}) or $(A,C)$ is detectable, the system
\begin{align}\label{e2}
\dot{x}_{i}(t)=Ax_{i}(t)
+c\sum_{j=1}^{N}l_{ij}P^{-1}C^{T}Cx_{j}(t)
\end{align}
can reach synchronization if the coupling strength $c$ is large enough.
\end{proposition}
\begin{proposition}
If $(A,B)$ is controllable, then
\begin{align}\label{e3}
\dot{x}_{i}(t)=A\delta x_{i}(t)
+cP^{-1}BB^{T}\sum_{j=1}^{N}l_{ij} x_{j}(t)
\end{align}
can reach consensus for sufficient large constant $c$.

Because in case $(A,B)$ is controllable, the QUAD condition
\begin{align}
(x-y)^{T}&P\bigg\{A(x-y)-P^{-1}BB^{T}[x- y]\bigg\}
\\&\le -\epsilon(x-y)^{T}(x-y)
\end{align}
is satisfied,
\end{proposition}

%and (\ref{e1}) takes
%\begin{align}\label{e2}
%\dot{e}_{i}(t)=Ae_{i}(t)
%+cP^{-1}C^{T}C\sum_{j=1}^{N}l_{ij}e_{j}(t)
%\end{align}
%or
%\begin{align}\label{e2}
%\dot{\delta} e_{i}(t)=A\delta e_{i}(t)
%+cP^{-1}C^{T}C\sum_{j=1}^{N}l_{ij}\delta e_{j}(t)
%\end{align}
%where $\delta e_{j}(t)=e_{j}(t)-\bar{e}(t)$, $\bar{e}(t)=\sum_{i=1}^{N}\xi_i e_{i}(t)$.

As direct consequences of Proposition 3, we have
\begin{proposition}
If $(A,B)$ is controllable, then
\begin{eqnarray}\label{pinc}
\left\{\begin{array}{cc}\frac{dx_1(t)}{dt}&=Ax_1(t)
+c\sum\limits_{j=1}^m l_{1j}P^{-1}BB^{T} x_j(t)\\
&-c\varepsilon(x_{1}(t)-s(t)),\\
\frac{dx_i(t)}{dt}&=Ax_i(t)+c\sum\limits_{j=1}^ml_{ij}P^{-1}BB^{T} x_j(t),\\
&i=2,\cdots,m\end{array}\right.
\end{eqnarray}
%where the matrix $P$ satisfies the QUAD condition
%\begin{align}
%(x-y)^{T}&P\bigg\{A(x-y)-P^{-1}BB^{T}[x- y]\bigg\}
%\\&\le -\epsilon(x-y)^{T}(x-y)
%\end{align}
can reach consensus to the trajectory $\dot{s}(t)=s(t)$ for sufficient large constant $c$.
\end{proposition}

In \cite{Li2,Li5}, fully distributed consensus protocols for linear
multi-agent systems were discussed.
In fact, we should consider following systems
\begin{eqnarray}
\left\{\begin{array}{cc}\frac{dx_1(t)}{dt}&=Ax_1(t)
+\sum\limits_{j=1}^m c_{i}(t)l_{1j}P^{-1}B^{T}Bc_{j}(t)x_j(t)\\
&-c_{i}(t)\varepsilon(x_{1}(t)-s(t)),\\
\frac{dx_i(t)}{dt}&=Ax_i(t)
+\sum\limits_{j=1}^ma_{ij}P^{-1}B^{T}Bc_{j}(t)x_j(t),\\
&i=2,\cdots,m\\
 &\dot{c}_{i}(t) =\frac{\alpha}{2}(x_i(t)-s(t))^{T}BB^{T}(x_i(t)-s(t))
\end{array}\right.
\label{pin2}
\end{eqnarray}
which can be rewritten as
\begin{eqnarray}
\left\{\begin{array}{cc}\frac{dx_i(t)}{dt}&=Ax_i(t)
+\sum\limits_{j=1}^m c_{i}(t)\tilde{l}_{1j}P^{-1}B^{T}Bc_{j}(t)x_j(t),\\
&i=1,\cdots,m\\
 &\dot{c}_{i}(t) =\frac{\alpha}{2}(x_i(t)-s(t))^{T}BB^{T}(x_i(t)-s(t))\end{array}\right. \label{pin2}
\end{eqnarray}
where $c(0)\geq 0$ and $\alpha>0$, can synchronize the coupled
system to the given trajectory $s(t)$.

\section{Comparisons}

In \cite{Chen1} 2006, time varying synchronization state $\bar{x}(t)$  as a non-orthogonal projection in synchronization manifold was first introduced and a distance between the state and the synchronization manifold $\delta(x(t)=x(t)-\bar{x}(t)$ was used to discuss synchronization was proposed and played key role.

It is clear that $\delta(t)$ used in \cite{Li1,Li2,Li3,Li4,Li5} and other papers, $\delta(t)$ is nothing new other than the $\delta(x(t))=x(t)-\bar{x}(t)$.

\subsection{{ On the paper Consensus of Multiagent Systems and Synchronization of Complex Networks: A Unified Viewpoint \cite{Li1}}}

Large number of results concerning with the consensus of multi-agents can be derived from above Theorem 1 and Theorem 2 as special cases.

In fact, let $f(x)=Ax$, synchronization model becomes consensus of multi-agents
\begin{align}
\dot{x}_{i}(t)=Ax_{i}(t)+c\sum_{j=1}^{m}l_{ij}\Gamma x_{j}(t)
\end{align}
and the local consensus and global consensus are equivalent. Then, as special cases of Theorem1 and Theorem 2, we have following two Theorems.
\begin{theorem}\quad
Let $\lambda_{2},\lambda_{3},\cdots,\lambda_{m}$ be the non-zero
eigenvalues of the coupling matrix $L$.
If either one condition is satisfied
\begin{enumerate}
\item
all variational equations
\begin{eqnarray}
\frac{d z(t)}{dt}=[A+\lambda_{k}\Gamma]z(t),\quad
k=2,3,\cdots,m
\end{eqnarray}
are exponentially stable,
\item
or there exist a positive definite matrix $P$ and a constant $\epsilon>0$, such that
\begin{eqnarray}
\bigg\{P(A+\lambda_{k}\Gamma)\bigg\}^{s}<-\epsilon I_{n},\quad
k=2,3,\cdots,m %\label{localsyn2}
\end{eqnarray}
\end{enumerate}
then the synchronization manifold
$\mathcal S$ is globally exponentially stable for the coupled system
\begin{align}
\dot{x}_{i}(t)=Ax+\sum_{j=1}^{m}l_{ij}\Gamma x_{j}(t)
\end{align}
\end{theorem}
\begin{theorem}\quad
Let $0>\lambda_{1}\ge\lambda_{2}\ge\cdots\ge\lambda_{m}$ be the
eigenvalues of the coupling matrix $\tilde{L}$. If either one condition is satisfied
\begin{enumerate}
\item
all variational
equations
\begin{eqnarray}
\frac{d z(t)}{dt}=[A+\lambda_{k}\Gamma]z(t),\quad
k=1,2,\cdots,m
\end{eqnarray}
are exponentially stable,
\item
or there exist a positive definite matrix $P$ and a constant $\epsilon>0$, such that
\begin{eqnarray}
\bigg\{P(A+\lambda_{k}\Gamma)\bigg\}^{s}<-\epsilon I_{n},\quad
k=1,2,\cdots,m %\label{localsyn2}
\end{eqnarray}
\end{enumerate}
then $s(t)$ satisfying $\dot{s}(t)=As(t)$ is globally exponentially stable for the coupled system
\begin{align}
\dot{x}_{i}(t)=Ax_{i}(t)+\sum_{j=1}^{m}\tilde{l}_{ij}\Gamma (x_{j}(t)-s(t))
\end{align}
\end{theorem}

We will show that the results given in the  following observer-type consensus protocol of multi-agents and synchronization of complex networks discussed in \cite{Li1,Li2} and some other papers
\begin{align}\label{Lia}
\left\{\begin{array}{ll}
\dot{x}_{i}(t)&=Ax_{i}(t)+u_{i}(t),~~y_{i}(t)=Cx_{i}(t),\\
\dot{v}_{i}(t)&=(A+BK)v_{i}(t)+cF\sum_{j=1}^{N}Cl_{ij}(v_{j}(t)-x_{j}(t))\\
 u_{i}&=Kv_{i}\end{array}\right.
\end{align}
can be easily derived from the results in \cite{Chen1}.

By routine technique used in linear system theory, let $e_{i}(t)=v_{i}(t)-x_{i}(t)$, then (\ref{Lia}) becomes
\begin{align}\label{Liaa}
\left\{\begin{array}{ll}
\dot{x}_{i}(t)&=Ax_{i}(t)+BKv_{i}(t),\\
\dot{e}_{i}(t)&=Ae_{i}(t)+cF\sum_{j=1}^{N}Cl_{ij}e_{j}(t)\end{array}\right.
\end{align}

In case $(A,B)$ is controllable, the synchronization of the system (\ref{Lia}) transfers to the synchronization of the system
\begin{align}%\label{e1}
\dot{e}_{i}(t)=Ae_{i}(t)
+cFC\sum_{j=1}^{N}l_{ij}e_{j}(t)
\end{align}
By previous results, we have
\begin{theorem}\quad
Let $\lambda_{2},\lambda_{3},\cdots,\lambda_{m}$ be the non-zero
eigenvalues of the coupling matrix $L$. If $(A,B)$ is controllable. either one condition is satisfied
\begin{enumerate}
\item
all variational
equations
\begin{eqnarray}
\frac{d z(t)}{dt}=[A+c\lambda_{k}FC]z(t),\quad
k=2,3,\cdots,m
\end{eqnarray}
are exponentially stable,
\item
or there exist a positive definite matrix $P$ and a constant $\epsilon>0$, such that
\begin{eqnarray}
\bigg\{P(A+c\lambda_{k}FC)\bigg\}^{s}<-\epsilon I_{n},\quad
k=2,\cdots,m %\label{localsyn2}
\end{eqnarray}
\end{enumerate}
then the synchronization manifold
$\mathcal S$ is globally exponentially stable for the coupled system (\ref{Lia})

\end{theorem}

\begin{theorem}\quad
Let $0>\lambda_{1}\ge\lambda_{2}\ge\cdots\ge\lambda_{m}$ be the
eigenvalues of the coupling matrix $\tilde{L}$. If either one condition is satisfied
\begin{enumerate}
\item
all variational
equations
\begin{eqnarray}
\frac{d z(t)}{dt}=[A+c\lambda_{k}FC]z(t),\quad
k=1,2,\cdots,m
\end{eqnarray}
are exponentially stable,
\item
or there exist a positive definite matrix $P$ and a constant $\epsilon>0$, such that
\begin{eqnarray}
\bigg\{P(A+c\lambda_{k}FC)\bigg\}^{s}<-\epsilon I_{n},\quad
k=1,2,\cdots,m %\label{localsyn2}
\end{eqnarray}
\end{enumerate}
then $0$ is exponentially stable for the coupled system
\begin{align}
\dot{x}_{i}(t)=Ax_{i}(t)+c\sum_{j=1}^{m}\tilde{l}_{ij}FC x_{j}(t)
\end{align}
\end{theorem}

Therefore, Theorem 1 in \cite{Li1} is just item 1 in Theorem 5 given above, and is a simple consequence of the results given in \cite{Chen1}.
%Moreover, the derivations presented in \cite{Li} are als
Lemma 1 in \cite{Zhang} is just item 1 in Theorem 6 given above, and is simple consequence of the results given in \cite{Chen1}.

In case $(A,C)$ is detectable, by (\ref{LMIC}), there exists a positive definite matrix $P$ such that
\begin{align}\label{de}
PA+A^{T}P-C^{T}C<-\epsilon I_{n}
\end{align}
%which was also reported in \cite{Li1}.

In this case, let $\Gamma=P^{-1}C^{T}C$, and by (\ref{de}), if $c>\frac{1}{|Re{\lambda_{2}}|}$, we have
\begin{eqnarray}
\bigg\{P(A+\lambda_{k}\Gamma)\bigg\}^{s}=PA+A^{T}P+cRe(\lambda_{k})C^{T}C<-\epsilon I_{n}
\end{eqnarray}

Therefore, by the item 2 in Theorem 3, we have
\begin{corollary}
If $(A,C)$ is detectable, and $c>\frac{1}{|Re{\lambda_{2}}|}$, the system
\begin{align}
\dot{x}_{i}(t)=Ax_{i}(t)+c\sum_{j=1}^{m}l_{ij}P^{-1}C^{T}C x_{j}(t)
\end{align}
is exponentially stable.
\end{corollary}

Many results given in \cite{Li2} can also be obtained as direct consequences of those given \cite{Chen1}.

\subsection{ Consensus of Multi-Agent Systems With General
Linear and Lipschitz Nonlinear Dynamics \cite{Li3}}

Similarly, in case $(A,B)$ is controllable, then there exists a positive definite matrix $P$ such that
\begin{align}
PA+A^{T}P-BB^{T}<-\epsilon I_{n}
\end{align}
Therefore, we have
\begin{corollary}
If $L$ is symmetric, $(A,B)$ is controllable, and $c>\frac{1}{|\lambda_{2}|}$, the system
\begin{align}
\dot{x}_{i}(t)=Ax_{i}(t)+c\sum_{j=1}^{m}l_{ij}P^{-1}BB^{T} x_{j}(t)
\end{align}
is exponentially stable.
\end{corollary}

It means that Lemma 1 in \cite{Li3} is a direct consequence of the Corollary given in \cite{Chen1}.

As pointed out above that the adaptive algorithm
\begin{align}
\left\{\begin{array}{ll}\frac{dx_i(t)}{dt}&=Ax_i(t))+c\sum\limits_{j=1}^m w_{ij}(t)\Gamma(x_j(t)-x_{i}(t)),\\
\dot{w}_{ij}(t) &=\rho_{ij}\xi_{i}
[x_i(t)-x_j(t)]^{T}\Gamma[x_i(t)-x_j(t)]
\end{array}\right.
\end{align}
for the system model
\begin{align}
\dot{x}_{i}(t)=Ax_i(t)+c\sum_{j=1}^{m}l_{ij}\Gamma x_{j}(t)
\end{align}
discussed in \cite{Li3,Li5}, where $L$ corresponds to an indirected graph, is a special case of the adaptive algorithm
\begin{align}
\left\{\begin{array}{ll}\frac{dx_i(t)}{dt}&=f(x_i(t)))+c\sum\limits_{j=1}^m w_{ij}(t)\Gamma(x_j(t)-x_{i}(t)),\\
\dot{w}_{ij}(t)& =\rho_{ij}\xi_{i}
[x_i(t)-\bar{x}(t)]^{T}\Gamma[x_i(t)-x_j(t)]
\end{array}\right. \label{pin3}
\end{align}
for the system
\begin{align}
\dot{x}_{i}(t)=f(x_i(t))+c\sum_{j=1}^{m}l_{ij}\Gamma x_{j}(t)
\end{align}
where $L$ corresponds to a direct graph discussed in \cite{Chen4} for adaptive cluster synchronization algorithm

Moreover, Theorem 1 in \cite{Li3} is a direct consequence of the corresponding adaptive cluster synchronization algorithm in \cite{Chen4}.

In \cite{Li3}, authors considered following mixed model
\begin{align}
\dot{x}_{i}(t)=Ax_{i}(t)+f(x_{i}(t))+\sum_{i=1}^{m}l_{ij}\Gamma x_{j}(t)
\end{align}

In fact, let $g(x)=Ax+f(x)$, which is a special case of the model (\ref{model1}).
\begin{align}
\dot{x}_{i}(t)=g(x_{i}(t))+\sum_{i=1}^{m} l_{ij}\Gamma x_{j}(t)
\end{align}
discussed in \cite{Chen1}.

Notice $(x-y)^{T}P[f(x)-f(y)]\le (x-y)^{T}(PP+\gamma^{2}I)(x-y)$, under the Theorem 2' assumptions, we have
\begin{align}
(x-y)^{T}&P\bigg\{A(x-y)+[f(x)-f(y)]-P^{-1}\Gamma [x- y]\bigg\}\nonumber\\&\le -\epsilon
(x-y)^{T}(x-y)
\end{align}
with $\Gamma=BB^{T}$, which is equivalent to
\begin{equation}
\left(\begin{array}{cc} PA+A^TP-BB^{T}-\gamma^{2}I & P\\ P &
-I\end{array}\right)>0 \label{LMI}
\end{equation}

Therefore, Theorem 2 in \cite{Li3} is a direct consequence of the results given in \cite{Chen1}.

\subsection{ Designing Fully Distributed Consensus Protocols for Linear
Multi-Agent Systems With Directed Graphs \cite{Li4}}

In \cite{Li4}, authors discussed leader-follower consensus problem for the agent. In fact, it is nothing new other than Pinning Complex Networks by a Single Controller discussed in
\cite{Chen2}.

In \cite{Li4}, following Lemma was given

Lemma 4 There exists a positive diagonal matrix G such that
$GL_{1} + L_{1}^{T}G>0$, where $G>0$.
One such G is given by $diag(q_1,\cdots, q_{N})$, where
$q=[q_{1},\cdots, q_{N}]^{T}=(L_{1}^{T})^{-1}1$.

In fact, it has been pointed out many years ago in \cite{Chen2}, where it was revealed  that all the eigenvalues of the matrix $\{\Xi \tilde{L}\}^{s}=\frac{1}{2}[\Xi \tilde{L}+\tilde{L}^{T}\Xi]$ are negative.

In \cite{Li2,Li5}, fully distributed consensus protocols for linear
multi-agent systems were discussed.
In fact, we should consider following systems
\begin{eqnarray}
\left\{\begin{array}{cc}\frac{dx_1(t)}{dt}&=Ax_1(t)
+\sum\limits_{j=1}^m c_{i}(t)l_{1j}P^{-1}B^{T}Bc_{j}(t)x_j(t)\\
&-c_{i}(t)\varepsilon(x_{1}(t)-s(t)),\\
\frac{dx_i(t)}{dt}&=Ax_i(t)
+\sum\limits_{j=1}^ma_{ij}P^{-1}B^{T}Bc_{j}(t)x_j(t),\\
 &\dot{c}_{i}(t) =\frac{\alpha}{2}\delta
x_i^T(t)BB^{T}\delta x_i(t)
\end{array}\right.
\label{pin2}
\end{eqnarray}
which can be rewritten as
\begin{eqnarray}
\left\{\begin{array}{cc}\frac{dx_i(t)}{dt}&=Ax_i(t)
+\sum\limits_{j=1}^m c_{i}(t)\tilde{l}_{1j}P^{-1}B^{T}Bc_{j}(t)x_j(t),\\
 &\dot{c}_{i}(t) =\frac{\alpha}{2}\delta
x_i^T(t)BB^{T}\delta x_i(t)\end{array}\right. \label{pin2}
\end{eqnarray}
where $c(0)\geq 0$ and $\alpha>0$, can synchronize the coupled
system to the given trajectory $s(t)$.

%In \cite{Chen5}, it was pointed out that many results given in \cite{Li1} can be obtained %from those given in \cite{Chen1}.

\section{Conclusions}

In this note, we revisit synchronization and consensus of multi-agents. As pointed out in \cite{Chen1}, synchronization relates two main points, one is connection structure, and the other is the intrinsic property of the uncoupled system.
\begin{itemize}
\item
In \cite{Chen1} 2006, time varying synchronization state $\bar{x}(t)$  as a non-orthogonal projection in synchronization manifold was first introduced and a distance between the state and the synchronization manifold $\delta(x(t))=x(t)-\bar{x}(t)$ was used to discuss synchronization was proposed and played key role. It describe the connection structure.
\item
It is clear that $\delta(t)$ used in \cite{Li1,Li2,Li3,Li4,Li5} and other papers, is nothing new other than the $\delta(x(t)=x(t)-\bar{x}(t)$, though the authors did not mention this fact and cited \cite{Chen1}.

\item
In \cite{Chen1} 2006, QUAD condition is introduced, which describes intrinsic property of the uncoupled system.

\item
Based on non-orthogonal projection, $\delta(x(t))$ and QUAD condition, conditions to ensure synchronization are given.

\item
It is clear that
\begin{align}\label{conclusion}
\frac{d x_{i}(t)}{dt}=Ax_{i}(t)
+c\sum\limits_{j=1}^{N}l_{ij}\Gamma x_{j}(t),\quad i=1,\cdots,N
\end{align}
is a special case of
\begin{align}\label{conclusion1}
\frac{d x_{i}(t)}{dt}=f(x_{i}(t))
+c\sum\limits_{j=1}^{N}l_{ij}\Gamma x_{j}(t),\quad i=1,\cdots,N
\end{align}
Therefore, all the results on synchronization model (\ref{conclusion1}) can apply to consensus of multi-agents model (\ref{conclusion}). All the results given in \cite{Li1,Li2,Li3,Li4,Li5} can be given as applications of the \cite{Chen1}.

\item
In fact, all papers on consensus focus on the QUAD condition
\begin{align}\label{conclusion1}
PA+A^{T}P-BB^{T}<-\epsilon I_{n}
\end{align}
As we point out that it is a natural consequence of the controllability, and is just another expression of QUAD condition.

\end{itemize}

%\end{CJK*}

\begin{thebibliography}{99}
\bibitem{Li1}
Zhongkui Li, Zhisheng Duan, Guanrong Chen, and Lin Huang
Consensus of Multiagent Systems and Synchronization of Complex Networks: A
Unified Viewpoint, {\it IEEE Trans. Circuits Syst. I}, vol. 57, pp. 213-224, 2010.

\bibitem{Li2}
Z. Li, Z. Duan, and G. Chen, ¡°Dynamic consensus of linear multi-agent
systems,¡± IET Control Theory and Applications, vol. 5, no. 1, pp. 19¨C28,
2011.

\bibitem{Li3}
Zhongkui Li, Wei Ren, Xiangdong Liu, and Mengyin Fu
Consensus of Multi-Agent Systems With General
Linear and Lipschitz Nonlinear Dynamics, IEEE TRANSACTIONS ON AUTOMATIC CONTROL, 60, 4, 1786-1791, 2013

\bibitem{Li4}
Zhongkui Li, Guanghui Wen,  Zhisheng Duan, and Wei Ren, Designing Fully Distributed Consensus Protocols for Linear Multi-Agent SystemsWith Directed Graphs Using Distributed Adaptive Protocols, IEEE
IEEE TRANSACTIONS ON AUTOMATIC CONTROL, 60, 4, 1152-1157, 2015

\bibitem{Li5}
Zhongkui Li, Wei Ren, Xiangdong Liu, Lihua Xie, Distributed consensus of linear multi-agent systems with adaptive dynamic protocols, Automatica 49 (2013) 1986-1995

\bibitem{Tuna}
S. Tuna, ¡°Conditions for synchronizability in arrays of coupled linear
systems,¡± IEEE Trans. Autom. Control, vol. 54, no. 10, pp. 2416¨C2420,
Oct. 2009.

\bibitem{Zhang}
H. Zhang, F. Lewis, and A. Das, ¡°Optimal design for synchronization of
cooperative systems: State feedback, observer, output feedback,¡± IEEE
Trans. Autom. Control, vol. 56, no. 8, pp. 1948¨C1952, 2011.

\bibitem{Chen1}
Wenlian Lu, Tianping Chen, ¡±New Approach to Synchronization Analysis
of Linearly Coupled Ordinary Differential Systems¡±, Physica D, 213,
2006, 214-230

\bibitem{book}
Tianping Chen and Wenlian Lu, "Theory of Cordination in Complex Networks"

\bibitem{Chen2}
Tianping Chen, Xiwei Liu, and Wenlian Lu, "Pinning Complex Networks by a Single Controller", IEEE Transactions on Circuits and Systems-I: Regular Papers, 54(6), 2007, 1317-1326

\bibitem{Chen3}
Wenlian Lu, Tianping Chen, "Synchronization  of Coupled Connected Neural Networks With Delays" IEEE Transactions on Circuits and Systems-I, Regular Papers, 51(12), (2004), 2491-2503

\bibitem{Chen4}
Wenlian Lu, Bo Liu, and Tianping Chen, Cluster synchronization in networks of coupled nonidentical dynamical systems, CHAOS 20, 013120,(2010)

\bibitem{Chen5}
Xiwei Liu and Tianping Chen, Cluster Synchronzation for Linearly
Coupled Complex Networks, Journal of Industrial and Management Optimization, 7(1), February 2011, 87-101

\bibitem{Chen6}
Tianping Chen, Synchronization, Consensus of Complex Networks
and their relationships, Arxiv: 2240 0762, 2018

\bibitem{Yu1}
Wenwu Yu Guanrong Chen, Jinhu L¨¹, On pinning synchronization of complex dynamical networks,
Automatica 45 (2009) 429-435
\bibitem{Yu2}
Wenwu Yu Guanrong Chen, Jinhu L¨¹, And Jurgen Kurths, Synchronization via pinning control on general complex networks, SIAM J. CONTROL OPTIM.
Vol. 51, No. 2, pp. 1395¨C1416
\end{thebibliography}
\end{document}